\documentclass[11pt,onecolumn,aps,superscriptaddress,amsmath,amssymb,
nofootinbib]{revtex4}      

\usepackage{graphicx}

\usepackage{graphicx,epsfig}

\usepackage{dcolumn}
\usepackage{bm}

\newcommand{\Z}{{\mathbb Z}}

\begin{document}

\begin{flushright}
\baselineskip=12pt \normalsize
\smallskip
\end{flushright}

\centerline{
\large \bf Natural Four-Generation Mass Textures in MSSM Brane Worlds}

\vspace{1cm}

\centerline{\sc Richard F. Lebed and Van E. Mayes}

\vspace{10mm}

\centerline{\it Department of Physics, Arizona State University}
\centerline{\it Tempe, AZ 85287-1504, USA}

\centerline{\bf Abstract}
A fourth generation of Standard Model (SM) fermions is usually
considered unlikely due to constraints from direct searches,
electroweak precision measurements, and perturbative unitarity.  We
show that fermion mass textures consistent with all constraints may be
obtained naturally in a model with four generations constructed from
intersecting D6 branes on a $T^6/(\Z_2 \times \Z_2)$ orientifold.  The
Yukawa matrices of the model are rank 2, so that only the third- and
fourth-generation fermions obtain masses at the trilinear level.  The
first two generations obtain masses via higher-order couplings and are
therefore naturally lighter.  In addition, we find that the third and
fourth generation automatically split in mass, but do not mix at
leading order.  Furthermore, the SM gauge couplings automatically
unify at the string scale, and all the hidden-sector gauge groups
become confining in the range $10^{13}$--$10^{16}$~GeV, so that the
model becomes effectively a four-generation MSSM at low energies.

\newpage

\section{Introduction}

Current high-energy experimental data supports just three generations
of chiral fermions.  However, the existence of a fourth generation
remains viable provided the mass of the fourth-generation neutrino
$\nu^{\, \prime}$ satisfies $m_{\nu^\prime} \! > \! \frac{1}{2} M_Z$,
and the corresponding charged fermion masses $m_{t^\prime}$,
$m_{b^\prime}$, and $m_{\tau^\prime}$ lie in the correct mass ranges
to avoid constraints from direct searches and precision electroweak
measurements~\cite{HolHouHurManSulUne09}.  Experimentally, various
anomalies indicate possible manifestations of a fourth generation.
Recent analysis from WMAP7~\cite {Wmap7} suggests a number of
relativistic neutrino flavors that is greater than three, $N_{\rm eff}
= 4.34 ^{+0.86}_{-0.88}$.  Explanations have been proposed based on
sterile neutrinos with masses at sub-eV scales~\cite{KraLunSmi10} (see
also Refs.~\cite{Khl99} for earlier discussion of astrophysical
implications of the fourth generation).  In addition, recent
observations from MiniBooNE~\cite{AguilarArevalo:2010wv} and the
observation of a reactor antineutrino anomaly~\cite{Mention:2011rk}
may indicate the existence of one or more light, sterile neutrinos.
Furthermore, the existence of a fourth generation generically
introduces additional sources of CP-violating
effects~\cite{Lunghi:2008aa,Soni:2008bc, Soni:2010xh}, which are
important for baryogenesis and could explain observed anomalies in
flavor physics~\cite{Lunghi:2010gv}.

Although the existence of a fourth generation is still viable, it has
long been considered unlikely.  In part, this predisposition relies
upon the perceived unnaturalness of the allowed fourth-generation
fermion masses, which are strongly constrained by the limits provided
by direct searches and precision electroweak measurements.  In
particular, these constraints mandate a large $\nu^\prime$ mass in
comparison to the tiny masses of the observed three neutrinos.
Moreover, the fourth-generation quark masses are constrained by the
perturbative unitarity of heavy-fermion scattering
amplitudes~\cite{Chanowitz:1978uj} to be $\lesssim 500$--600~GeV\@.
Thus, the $t^\prime$ quark mass $m_{t'}$ can only lie between $\sim$
2--3 times the $m_t$, a small difference when compared to the mass
hierarchy $m_t/m_c \! \approx \! 135$.  Indeed, for the known families
of quarks and leptons, progressively larger mass splittings between
each sequential family appears to be the rule, a pattern that cannot
hold in four-generation models due to the limited allowed range for
$m_{t'}$.  Furthermore, contributions to the oblique $S$ and $T$
parameters from an additional chiral fermion generation are
intolerably large unless the masses of the $t'$ and $b'$ quarks are
equal to within $m_{t'} \! - m_{b'} \simeq
(1+\frac{1}{5}\mbox{ln}\frac{m_h}{115})\times 50$~GeV, where $m_h$ is
the Higgs mass~\cite{Kribs:2007nz}.  Moreover, mixing between the
third- and fourth-generation quarks is often presumed small to
accommodate the unitarity of the CKM matrix.  These features of
four-family models appear to require significant fine tuning to
explain, and therefore a fourth generation of chiral fermions is
usually deemed improbable.

In fact, the pioneering work of Ref.~\cite{Kribs:2007nz} showed that
fourth-generation fermions are phenomenologically perfectly viable,
apart from mild tuning of the masses within the fermion SU(2)$_L$
doublets.  The direct search limits quoted therein are $m_{\nu^{\,
\prime}}, m_{\tau^\prime} > 100$~GeV and $m_{t^\prime}, m_{b^\prime} >
258$~GeV, and the electroweak precision-based constraints are given by
$m_{\tau^\prime} \! - m_{\nu^{\, \prime}} \approx$~30--60~GeV and the
$m_{t'} \! - m_{b'}$ splitting mentioned above, while the most
stringent quark-mixing bounds are actually those with the first
generation, $|V_{ub'}|$, $|V_{t'd}| \lesssim 0.04$.  Subsequent bounds
from both observation and more detailed calculations ({\it
e.g.},~\cite{Hung:2007ak,Lenz}) modify these values slightly, but the
possibility of a fourth chiral fermion generation remains vibrant and
awaits the decisive verdict of the Large Hadron Collider.

Currently, string theory is the most promising approach for a
unification of gravity with quantum mechanics.  As such, it should
provide a first-principles explanation of the detailed properties of
our universe.  However, as is well known, string theory exhibits a
huge degeneracy of vacua, resulting in the so-called string landscape.
For many string compactifications, replication of chiral matter seems
to be a generic feature.  At present, the principle that sets the
precise number of observed chiral generations remains unknown.
Indeed, many consistent string vacua can be constructed that closely
resemble the Minimal Supersymmetric Standard Model (MSSM) but contain
different numbers of generations~\cite{Belitsky:2010zr}.

One particularly promising class of string compactifications employs D
branes on orientifolds (for reviews, see~\cite{Ura03,BluCveLanShi05,
BluKorLusSti06,Mar07}).  In such models chiral fermions---an intrinsic
feature of the Standard Model (SM)---arise from configurations with D
branes located at transversal orbifold/conifold
singularities~\cite{DouMoo96} and strings stretching between D branes
intersecting at angles~\cite{BerDouLei96,AldFraIbaRabUra00} (or, in
its T-dual picture, with magnetized
D~branes~\cite{Bac95,BluGorKorLus00,AngAntDudSag00}).  Within the
framework of D-brane model building on toroidal orientifolds, the
worldsheet areas $A_{abc}$ spanning the region bounded by D branes
(labeled by $a$, $b$, $c$) that support fermions and Higgses at their
intersections give rise to Yukawa couplings $Y_{abc}
\sim \exp(- A_{abc})$~\cite{AldFraIbaRabUra00,Cremades:2003qj}.  This
pattern naturally encodes the hierarchy of Yukawa couplings.  Indeed,
models exist in the literature in which realistic mass matrices and
CKM mixings for both three- and four- generation models may be
obtained~\cite{Belitsky:2010zr}.  While accommodating full rank-3 or
rank-4 Yukawa matrices is possible, one may nevertheless question
whether the degree of fine-tuning of parameters required is too large
to provide a completely natural explanation of the observed fermion
mass and mixing hierarchy.

In this paper we construct a four-generation MSSM with intersecting D6
branes on a $T^6/(\Z_2 \times \Z_2)$ orientifold.  We show that the
Yukawa matrices are rank 2, and thus only the heavier two generations
receive masses at the trilinear level, while the lighter two
generations can receive masses from higher-order
couplings.\footnote{In some cases, use of the term ``naturally
lighter'' must be carefully considered: Note, for example, the second-
to third-generation mass ratio $m_c/m_\tau \approx 0.7$. However, a
strong hierarchy remains when one compares only fermions of the same
type, {\it e.g.}, $m_\mu
\! \ll m_\tau$.}  Moreover, the third- and fourth-generation fermions
automatically split in mass, but we find that they do not mix at the
trilinear level and therefore do not generate $O(1)$ mixing angles.
Thus, one can obtain fermion mass textures naturally consistent with
fourth-generation constraints.  Finally, the tree-level gauge
couplings automatically unify at the string scale, and the
hidden-sector gauge groups introduced to satisfy tadpole constraints
become confining in the range 10$^{13}$--10$^{16}$~GeV, leaving only
the MSSM at low energies.

\section{A Four-Family MSSM}

\subsection{Model Building on $T^6/(\Z_2 \times \Z_2)$}

Let us begin by considering Type IIA string theory compactified on a
$T^6/(\Z_2\times \Z_2)$ orientifold~\cite{CveShiUra01a,CveShiUra01b},
where the six-torus is factorizable in terms of two-tori as $T^6 = T^2
\times T^2 \times T^2$, and the $i^{\rm th}$ two-torus, $i=1,2,3$,
has complex coordinates $z_i$.  The $\theta$ and $\omega$ generators
for the orbifold group $\Z_{2} \times \Z_{2}$ act on the complex
coordinates of $T^6$ as
\begin{eqnarray}
& \theta: & (z_1,z_2,z_3) \to (-z_1,-z_2,z_3)~,~ \nonumber \\ &
\omega: & (z_1,z_2,z_3) \to (z_1,-z_2,-z_3)~.~\,
\label{orbifold}
\end{eqnarray}
The orientifold projection is implemented by gauging the symmetry
$\Omega R$, where $\Omega$ is worldsheet parity, and $R$ is given by
conjugation:
\begin{eqnarray}
R: (z_1,z_2,z_3) \to ({\overline z}_1,{\overline z}_2,{\overline
z}_3)~.~\,    \label{orientifold}
\end{eqnarray}
As a result, one finds four kinds of orientifold 6-planes (O6 planes),
corresponding to the actions $\Omega R$, $\Omega R\theta$, $\Omega R
\omega$, and $\Omega R\theta\omega$.  Only two kinds of complex
structures are consistent with orientifold projection for a two-torus:
rectangular and tilted~\cite{CveShiUra01a,CveShiUra01b}, with $\beta_i
\! = \! 0, 1$ for an untilted or tilted $i^{\rm th}$ torus,
respectively.  In the following, we only consider rectangular two-tori
(but retain parameters allowing for tilt to permit the generalization
of the model presented here).  The homology three-cycles for a stack
$a$ of D6 branes and its orientifold image $a'$ are given by
\begin{equation}
[\Pi_a]=\prod_{i=1}^{3}(n^i_a[a_i]+ 2^{-\beta_i} l^i_a[b_i]),\;\;\;
[\Pi_{a'}]=\prod_{i=1}^{3}(n^i_a[a_i]- 2^{-\beta_i} l^i_a[b_i]) \, .
\end{equation}

The Ramond-Ramond (R-R) tadpole cancellation requires the total
homology-cycle \\ charge of D6 branes and O6 planes to
vanish~\cite{BluGorKorLus00}, which may be as expressed as
\begin{equation}
\sum_a N_a[\Pi_a]+\sum_a N_a[\Pi_{a'}]-4[\Pi_{O6}]=0 \, .
\end{equation}
It is convenient to use the parameters
\begin{equation}
\begin{array}{llll}
\tilde{A}_a = -l^1_a l^2_a l^3_a \, , & \tilde{B}_a =
l^1_a n^2_a n^3_a \, , & \tilde{C}_a = n^1_a l^2_a n^3_a \, , &
\tilde{D}_a = n^1_a n^2_a l^3_a \, , \\
A_a = -n^1_a n^2_a n^3_a \, , & B_a = n^1_a l^2_a l^3_a \, , & 
C_a = l^1_a n^2_a l^3_a \, , & D_a = l^1_a l^2_a l^3_a \, ,
\end{array}
\end{equation}
which were first introduced in~\cite{CvePapShi02}.  The tadpole
conditions can then be expressed as
\begin{equation}
\sum_a N_a A_a=\sum_a N_a B_a=\sum_a N_a C_a=\sum_a N_a D_a = -16 \, .
\label{tadpole}
\end{equation}

Completely eliminating the R-R tadpoles actually requires the
cancellation of D-brane charges as classified by K theory, which for
tori of arbitrary tilt for the $\Z_2 \times \Z_2$ orientifold
requires~\cite{BluCveLanShi05,Marchesano:2004xz,Chen:2005mm}
\begin{equation}
\sum_a N_a \tilde A_a \in 4 \Z \, , \ \
\sum_a N_a \tilde B_a \in 4 \Z \, , \ \
\sum_a N_a \tilde C_a \in 4 \Z \, , \ \
\sum_a N_a \tilde D_a \in 4 \Z \, .
\end{equation}

In order to preserve $\mathcal{N}=1$ supersymmetry in four dimensions,
the rotation angle of any D brane with respect to the orientifold
plane must be an element of SU(3)~\cite{BerDouLei96,CveShiUra01b}.
Consider the angles $\theta^i_a$ between each brane $a$ and the
$R$-invariant axis of the $i^{\mathrm{th}}$ torus; one requires that
$\theta^1_a + \theta^2_a +
\theta^3_a = 0$ mod $2\pi$, which means $\sin(\theta^1_a + \theta^2_a
+ \theta^3_a)= 0$ and $\cos(\theta^1_a +
\theta^2_a + \theta^3_a) > 0 $.  Define
\begin{equation}
\tan\theta^i_a= 2^{-\beta_i} \frac{l^i_a R^i_2}{n^i_a R^i_1} =
\frac{2^{-\beta_i} l^i_a}{n^i_a}\chi^i \, ,
\end{equation}
where $R^i_{1,2}$ are the radii of the $i^{\mathrm{th}}$ torus and
$\chi_i = R_2^i/R_1^i$ are the complex-structure moduli.  The
supersymmetry conditions can then be recast as~\cite{CvePapShi02}
\begin{eqnarray}
x_A\tilde{A_a}+x_B\tilde{B_a}+x_C\tilde{C_a}+x_D\tilde{D_a}=0 \, ,
\nonumber \\
A_a/x_A + B_a/x_B + C_a/x_C + D_a/x_D <0 \, ,
\end{eqnarray}
where $x_A$, $x_B$, $x_C$, $x_D$ are complex-structure parameters,
which are given by
\begin{equation}
x_A=\lambda,\;\;x_B=\lambda \cdot 2^{\beta_2+\beta_3}/\chi_2\chi_3,
\;\;x_C=\lambda \cdot 2^{\beta_1+\beta_3}/\chi_1\chi_3,\;\;
x_D=\lambda \cdot 2^{\beta_1+\beta_2}/\chi_1\chi_2 \, ,
\end{equation}
where $\lambda$ is a positive parameter introduced~\cite{CvePapShi02}
to put all the variables $A,B,C,D$ on an equal footing.  However,
among the $x_i$, only three are independent.

The initial U($N_a$) gauge group supported by a stack of $N_a$
identical D6 branes is broken down by the $\Z_2\times \Z_2$ symmetry
to a subgroup U($N_a/2$)~\cite{CveShiUra01b}.  Chiral matter fields
are formed from open strings whose two ends attach to different
stacks.  By using the Grassmann algebra
$[a_i][b_j]=-[b_j][a_i]=\delta_{ij}$ and $[a_i][a_j]=-[b_j][b_i]=0$,
one can calculate the intersection numbers between stacks $a$ and $b$
and obtain the multiplicity (${\cal M}$) of the corresponding
bifundamental representation:
\begin{equation}
{\cal M}\left(\frac{N_a}{2}, \frac{\overline{N_b}}{2}\right)
=I_{ab}=[\Pi_a][\Pi_b]= 2^{-k} \prod_{i=1}^3(n^i_a l^i_b - n^i_b
l^i_a) \, ,
\end{equation}
where $k \! \equiv \! \beta_1 \! + \! \beta_2 \! + \! \beta_3$ is the
total number of tilted tori.  Likewise, stack $a$ paired with the
orientifold image $b'$ of $b$ yields
\begin{equation}
{\cal M}\left(\frac{N_a}{2}, \frac{N_b}{2}\right)
=I_{ab'}=[\Pi_a][\Pi_{b'}]=-2^{-k} \prod_{i=1}^3(n^i_a l^i_b + n^i_b
l^i_a)
\, .
\end{equation}

Strings between a brane in stack $a$ and its orientifold image $a'$
yield chiral matter in the antisymmetric $a_A$ and symmetric $a_S$
representations of the group U($N_a/2$), with multiplicities
\begin{equation}
{\cal M}[(a_{A,a})_L]=\frac{1}{2}I_{aO6},\;\; {\cal M}[
(a_{A,a}+ a_{S,a})_L]=\frac{1}{2}
\left(I_{aa'}-\frac{1}{2}I_{aO6}\right) \, ,
\label{AandSreps}
\end{equation}
so that the net numbers of antisymmetric and symmetric representations
are given by
\begin{eqnarray}
{\cal M}(a_{A, \, a})=\frac{1}{2}
\left(I_{aa'}+\frac{1}{2}I_{aO6} \right) =
- 2^{1-k} [(2A_a-1)\tilde{A_a}-\tilde{B_a}-\tilde{C_a}-\tilde{D_a}] \,
, \nonumber \\ {\cal M}(a_{S, \, a})=\frac{1}{2}
\left( I_{aa'}-\frac{1}{2}I_{aO6} \right) =
- 2^{1-k} [(2A_a+1)\tilde{A_a}+\tilde{B_a}+\tilde{C_a}+\tilde{D_a}] \,
,
\label{netmult}
\end{eqnarray}
where
\begin{equation}
I_{aa'}=[\Pi_a][\Pi_{a'}]=-2^{3-k} \prod_{i=1}^3 \,n^i_a l^i_a \, ,
\end{equation}
\begin{equation}
I_{aO6}=[\Pi_a][\Pi_{O6}]= 2^{3-k}
(\tilde{A_a}+\tilde{B_a}+\tilde{C_a}+
\tilde{D_a}) \, .
\end{equation}
Note that the expressions Eqs.~(\ref{AandSreps}) and (\ref{netmult})
indicate the presence of non-chiral pairs of matter in the
antisymmetric representation that are masked by Eq.~(\ref{netmult}),
which gives only the net number.

A zero intersection number between two branes implies that the branes
are parallel on at least one torus.  At such types of intersection,
additional non-chiral (vectorlike) multiplet pairs from $ab+ba$,
$ab'+b'a$, and $aa'+a'a$ can
arise~\cite{PAUL}\footnote{Representations $({\rm
Anti}_a+\overline{\rm Anti}_a)$ occur at intersection of $a$ with
$a^{\prime}$ if they are parallel on at least one torus.}.  The
multiplicity of these non-chiral multiplet pairs is given by the
remainder of the intersection product, neglecting the null sector.
For example, if $(n^1_a l^1_b - n^1_b l^1_a)=0$ in $
I_{ab}=[\Pi_a][\Pi_b]=2^{-k}\prod_{i=1}^3(n^i_a l^i_b - n^i_b l^i_a)
$, then
\begin{equation}
{\cal M}\left[\left(\frac{N_a}{2},\frac{\overline{N_b}}{2}\right)
+\left(\frac{\overline{N_a}}{2},\frac{N_b}{2}\right)\right]
=\prod_{i=2}^3(n^i_a l^i_b - n^i_b l^i_a) \, .
\end{equation}
This result is useful since one can fill the spectrum with this matter
without affecting the required global conditions, because the total
effect of the pairs is zero. Typically in this type of model, the
Higgs fields arises from this non-chiral matter.  However, as we shall
see for the model constructed here, the Higgs fields appear in the
chiral sector of the model.

The total non-Abelian anomaly in intersecting brane-world models
cancels automatically when the R-R tadpole conditions are satisfied,
but additional mixed anomalies may be present.  For example, the mixed
gravitational anomalies that generate massive fields are not trivially
zero~\cite{CveShiUra01b}.  These anomalies are cancelled by a
generalized Green-Schwarz (G-S) mechanism that involves untwisted R-R
forms.  The couplings of the four untwisted R-R forms $B^i_2$ to the
U(1) field strength $F_a$ of each stack $a$ are given by
\begin{eqnarray}
 N_a l^1_a n^2_a n^3_a \int_{M4}B^1_2\wedge \textrm{tr}F_a,  \;\;
 N_a n^1_a l^2_a n^3_a \int_{M4}B^2_2\wedge \textrm{tr}F_a \, ,
  \nonumber \\
 N_a n^1_a n^2_a l^3_a \int_{M4}B^3_2\wedge \textrm{tr}F_a,  \;\;
-N_a l^1_a l^2_a l^3_a \int_{M4}B^4_2\wedge \textrm{tr}F_a \, .
\end{eqnarray}

These couplings then determine the exact linear combinations of U(1)
gauge bosons that acquire string-scale masses via the G-S mechanism.
If U(1)$_X$ is a linear combination of the U(1)s from each stack,
\begin{equation}
{\rm U}(1)_X \equiv \sum_a C_a {\rm U}(1)_a \, ,
\end{equation}
then the corresponding field strength must be orthogonal to those that
acquire G-S mass.  Thus, if a linear combination U(1)$_X$ satisfies
\begin{eqnarray}
\sum_a C_a N_a \tilde{B_a} =0, \;\; \sum_a C_a N_a \tilde{C_a} =0 \, ,
  \nonumber \\
\sum_a C_a N_a \tilde{D_a} =0, \;\; \sum_a C_a N_a \tilde{A_a} =0 \, ,
\label{GSeq}
\end{eqnarray}
the gauge boson of U(1)$_X$ acquires no G-S mass and is anomaly-free.

\subsection{The Model}

We present the D6-brane configurations, intersection numbers, and
complex-structure parameters of the model in Table~\ref{MI-Numbers},
and the resulting spectrum in Table~\ref{Spectrum}.  With this
configuration, all R-R tadpoles are cancelled, $\mathcal{N}=1$
supersymmetry is preserved, and K-theory conditions are satisfied.
The resulting model has gauge symmetry $[{\rm U}(4)_C \times {\rm
U}(2)_L \times {\rm U}(2)_R]_{\rm observable}\times [ {\rm
USp}(8)^3]_{\rm hidden}$, a left-right symmetric Pati-Salam symmetry
that provides (as seen below) a convenient origin for the SM
hypercharge U(1)$_Y$.

In this model the anomalies from the three global U(1)s of U(4)$_C$,
U(2)$_L$, and U(2)$_R$ are each cancelled by the G-S mechanism, and
the gauge fields of these U(1)s obtain masses via the linear $B\wedge
F$ couplings, except for one linear combination ${\rm U}(1)_X = {\rm
U}(1)_C + 2[{\rm U}(1)_{2L} + {\rm U}(1)_{2R}]$.  The effective gauge
symmetry of the observable sector is then ${\rm SU}(4)_C\times {\rm
SU}(2)_L\times {\rm SU}(2)_R\times {\rm U}(1)_X$.

The Pati-Salam gauge symmetry must then be broken to the SM\@.  The
breaking is accomplished by brane splitting, which corresponds in the
field theory description to giving VEVs along D-flat and F-flat
directions to some of the adjoint scalars that arise via open-string
moduli associated with each stack of D branes~\cite{Cvetic:2004nk}.
The $a$ stack of D6 branes may be split on any of the three two-tori
into $a1$ (quark) and $a2$ (lepton) stacks with $N_{a1}=6$ and
$N_{a2}=2$ D6 branes, respectively.  In same manner, the $c$ stack of
D6 branes is split into stacks $c1$ ($I_{3L} = +\frac 1 2$ fermions)
and $c2$ ($I_{3L} = -\frac 1 2$ fermions) such that $N_{c1} = 2$ and
$N_{c2} = 2$, respectively.  After the splitting, the observable gauge
symmetry is broken to ${\rm SU}(3)_C\times {\rm SU}(2)_L\times {\rm
U}(1)_{I_{3R}}\times {\rm U}(1)_{B-L} \times {\rm U}(1)_X$, where the
gauge bosons of ${\rm U}(1)_{I_{3R}}$, ${\rm U}(1)_{B-L}$, and ${\rm
U}(1)_X$ remain massless.  The ${\rm U}(1)_{I_{3R}}\times {\rm
U}(1)_{B-L}\times {\rm U}(1)_X$ gauge symmetry may then be broken
spontaneously if, for example, vectorlike particles with the quantum
numbers $({\bf 1}, {\bf 1}, \frac 1 2, -1, 3)$ and $({\bf 1}, {\bf 1},
-\frac 1 2, 1, -3)$ under the ${\rm SU}(3)_C\times {\rm SU}(2)_L\times
{\rm U}(1)_{I_{3R}} \times {\rm U}(1)_{B-L} \times {\rm U}(1)_X$ gauge
symmetry from the $a_2 c_2'$ intersections ($\Phi$, $\bar \Phi$ in
Table~\ref{Spectrum}) receive VEVs,\footnote{The $a_2 c_1'$
intersections were incorrectly indicated for this purpose in
Ref.~\cite{Belitsky:2010zr}.} leaving only the single U(1)
corresponding to SM hypercharge,
%
\begin{equation}
{\rm U}(1)_Y = \frac{1}{6}[{\rm U}(1)_{a1} - 3{\rm U}(1)_{a2} - 3{\rm
U}(1)_{c1} + 3{\rm U}(1)_{c2}]
\, ,
\end{equation}
and the overall gauge group ${\rm SU}(3)_C\times {\rm SU}(2)_L\times
{\rm U}(1)_Y \times {\rm USp}(8)^3$.

The observable sector of the model then possesses the gauge symmetry
and matter content of a four-generation MSSM with an extended Higgs
sector.  Note that the four pairs of Higgs fields in this model appear
in the chiral spectrum, rather than arising as vectorlike matter from
$\mathcal{N}=2$ subsectors, as is the case for the models studied
in~\cite{CheLiMayNan07}, as well as most models in the literature.
The extra matter in this model includes fields charged under the
hidden-sector gauge groups, vectorlike matter between nonintersecting
pairs of branes, and the chiral adjoints associated with each stack of
branes.  In addition, matter occurs in the symmetric triplet and
antisymmetric singlet representations of SU(2)$_L$.
 
\begin{table}[t]
\footnotesize
\renewcommand{\arraystretch}{1.0}
\caption{D6-brane configurations and intersection numbers for
a four-family Pati-Salam model on a Type-IIA $T^6 / (\Z_2
\times \Z_2)$ orientifold, with no tilted tori.  The complete gauge
symmetry is $[{\rm U}(4)_C \times {\rm U}(2)_L \times {\rm
U}(2)_R]_{\rm observable} \times [ {\rm USp}(8)^3]_{\rm hidden}$, and
the complex-structure parameters are $\chi_1=2\chi_2=2\chi_3$.  The
parameters $\beta_i$ are the $\beta$-function coefficients for the
${\rm USp(8)}_i$ gauge groups.}
\label{MI-Numbers}
\begin{center}
\begin{tabular}{|c||c|c|c|c|c|c|c|c|c|c|c|}
\hline
& \multicolumn{11}{c|}{${\rm U}(4)_C\times {\rm U}(2)_L\times {\rm
U}(2)_R \times {\rm USp}(8)^3$}\\
\hline \hline  & $N$ & $(n^1,m^1)\times (n^2,m^2)\times

(n^3,m^3)$ & $n_{S}$& $n_{A}$ & $b$ & $b'$ & $c$ & $c'$& 1 & 2 & 3 \\

\hline

     $a$&  8& $(0,-1)\times (1,1)\times (1,1)$ & \ 0 & \ 0  & 4 & 0 &
$-4$ & 0 & 1 & $-1$ & \ 0 \\

    $b$&  4& $(2,1)\times (1,-1)\times (1,0)$ & \ 2 & $-2$  & - & - &
\ 4 & 0 & 0 & \ 1 & $-2$ \\

    $c$&  4& $(2,-1)\times (0,1)\times (1,-1)$ & $-2$ & \ 2  & - & - &
- & - & $-1$ & 0 & \ 2 \\

\hline

    1&   8& $(1,0)\times (1,0)\times (1,0)$ & \multicolumn{9}{c|}
{$\chi_1=2,~\chi_2=1,~\chi_3=1$}\\

    2&   8& $(1,0)\times (0,-1)\times (0,1)$ & \multicolumn{9}{c|}
{$\beta^g_1=-3,~\beta^g_2=-3$}\\

    3&   8& $(0,-1)\times (1,0)\times (0,1)$& \multicolumn{9}{c|}
{$\beta^g_3=-2$}\\

\hline

\end{tabular}

\end{center}

\end{table}

\begin{table}
[htb] \footnotesize
\renewcommand{\arraystretch}{1.0}
\caption{The chiral and vectorlike superfields, their multiplicities
and quantum numbers under the gauge symmetry ${\rm SU}(4)_C\times {\rm
SU}(2)_L\times {\rm SU}(2)_R \times {\rm USp}(8)_1 \times
{\rm USp}(8)_2 \times {\rm USp}(8)_3$.}
\label{Spectrum}
\begin{center}
\begin{tabular}{|c||c|c||c|c|c||c|c|c|}\hline
& Mult. & Quantum Number & $Q_4$ & $Q_{2L}$ & $Q_{2R}$ & Field
\\
\hline\hline
$ab$ & 4 & $(4,\overline{2},1,1,1,1)$ & \ 1 & $-1$ & \ 0  &
$F_L(Q_L, L_L)$\\
$ac$ & 4 & $(\overline{4},1,2,1,1,1)$ & $-1$ & 0 & \ 1   &
$F_R(Q_R, L_R)$\\
$bc$ & 4 & $(1,2,\overline{2},1,1,1)$ & 0 & 1 & $-1$   &
$H_u^i$, $H_d^i$\\
\hline
$a1$ & 1 & $(4,1,1,\overline{8},1,1)$ & \ 1 & 0 & 0  & $X_{a1}$ \\
$a2$ & 1 & $(\overline{4},1,1,1,8,1)$ & $-1$ & 0 & 0   & $X_{a2}$
\\
$b2$ & 1 & $(1,2,1,1,\overline{8},1)$ & 0 & \ 1 & 0   & $X_{b2}$ \\
$b3$ & 2 & $(1,\overline{2},1,1,1,8)$ & 0 & $-1$ & 0    &
$X_{b3}^i$ \\
$c1$ & 1 & $(1,1,\overline{2},8,1,1)$ & 0 & 0 & $-1$    & $X_{c1}$
\\
$c3$ & 2 & $(1,1,2,1,1,\overline{8})$ & 0 & 0 & \ 1   &  $X_{c3}^i$
\\
$b_{S}$ & 2 & $(1,3,1,1,1,1)$ & 0 & \ 2 & 0   &  $T_L^i$ \\
$b_{A}$ & 2 & $(1,\overline{1},1,1,1,1)$ & 0 & $-2$ & 0   & $S_L^i$
\\
$c_{S}$ & 2 & $(1,1,\overline{3},1,1,1)$ & 0 & 0 & $-2$   & $T_R^i$
\\
$c_{A}$ & 2 & $(1,1,1,1,1,1)$ & 0 & 0 & \ 2   & $S_R^i$ \\
\hline\hline
$ab'$ & 2 & $(4,2,1,1,1,1)$ & \ 1 & \ 1 & 0  & \\
& 2 & $(\overline{4},\overline{2},1,1,1,1)$ & $-1$ & $-1$ & 0  &
\\
\hline
$ac'$ & 2 & $(4,1,2,1,1,1)$ & \ 1 & 0 & \ 1  & $\Phi_i$ \\
& 2 & $(\overline{4}, 1, \overline{2},1,1,1)$ & $-1$ & 0 & $-1$
& $\overline{\Phi}_i$\\
\hline
$bb'$ & 2 & $(1,\overline{1},1,1,1,1)$ & 0 & $-2$ & 0   & $s_L^i$ \\
      & 2 & $(1,1,1,1,1,1)$            & 0  & $ 2$ & 0  & $\bar{s}_L^i$ \\
\hline
$cc'$ & 2 & $(1,1,1,1,1,1)$ & 0 & $0$ & $2$   & $s_R^i$ \\
      & 2 & $(1,1,\overline{1},1,1,1)$ & 0 & 0  & $-2$  & $\bar{s}_R^i$ \\
\hline
$bc'$ & 1 & $(1,2,2,1,1,1)$ & 0 & \ 1 & \ 1  & $ H'$ \\
& 1 & $(1, \overline{2}, \overline{2},1,1,1)$ & 0 & $-1$ & $-1$ &
$\overline{H}'$\\
\hline
\end{tabular}
\end{center}
\end{table}

In order for the model to reproduce just the MSSM at low energies, the
gauge couplings must unify, and all extra matter besides the MSSM
states must become massive at high-energy scales.  Furthermore, since
the MSSM contains just one pair $H_{u,d}$ of light Higgs doublets, the
Higgs potential must be fine tuned.  Since the four Higgs fields in
this model appear in the chiral sector rather than as $\mathcal{N}=2$
vectorlike multiplets, no intrinsic $\mu$ parameter (whose real part
corresponds geometrically to the separation between stacks $b$ and
$c$) occurs since the $b$ and $c$ stacks intersect on all three
two-tori.  A single pair of light Higgs doublets may be obtained by
fine-tuning the $\mu$ term, which may be generated in the
superpotential $W$ via the higher-dimensional operators
\begin{eqnarray}
W & \supset & \frac{y^{ijkl}_{\mu}}{M_{\rm St}} S_L^i S_R^j
H_u^k H_d^l  ~,~\,
\label{eqn:HiggsSup}
\end{eqnarray}
where $y^{ijkl}_{\mu}$ are Yukawa couplings, and $M_{\rm St}$ is the
string scale.  In this scenario, the singlets $S^i_R$ may obtain
string or GUT-scale VEVs while preserving the D-flatness of U(1)$_R$,
and the singlets $S^i_L$ may obtain TeV-scale VEVs while preserving
the D-flatness of U(1)$_L$ (and hence are tied to electroweak symmetry
breaking).  The precise linear combinations that produce the two light
Higgs eigenstates $H_{u,d}$ obtained by fine-tuning the Higgs
potential via Eq.~(\ref{eqn:HiggsSup}) are then correlated with the
pattern of Higgs VEVs necessary to obtain Yukawa matrices for the
quarks and leptons,
\begin{equation}
H_{u,d} = \sum_i \frac{v^i_{u,d} \, H_{u,d}^i}
{\sqrt{\sum_j (v^j_{u,d})^2}},
\end{equation}
where $v^i_{u,d} = \left\langle H^i_{u,d} \right\rangle$.
 
The gauge coupling constant associated with a stack $P$ is given by
\begin{eqnarray}
g_{D6_P}^{-2} &=& |\mathrm{Re}\,(f_P)| \, , \label{idb:eq:gkf}
\end{eqnarray}
where $f_P$ is the holomorphic kinetic gauge function associated with
stack $P$, given by~\cite{BluKorLusSti06,CreIbaMar02A}
\begin{eqnarray}
f_P &=&
\frac{1}{4\kappa_P} \left( n_P^1\,n_P^2\,n_P^3\,s
- n_P^1\,l_P^2\,l_P^3\,u^1 - n_P^2\,l_P^1\,l_P^3\,u^2
- n_P^3\,l_P^1\,l_P^2\,u^3 \right) \, ,
\label{kingaugefun}
\end{eqnarray}
with $\kappa_P = 1$ for SU($N_P$) and $\kappa_P = 2$ for USp($2N_P$)
or SO($2N_P$) gauge groups, and $s$ and $u$ are moduli in the
supergravity basis.  The holomorphic gauge kinetic function associated
with the SM hypercharge U(1)$_Y$ is then given for this model by the
combination~\cite{BluLusSti03}
\begin{equation}
f_Y = \frac{1}{6}f_{a1} + \frac{1}{2}f_{a2} + \frac{1}{2}f_{c1} +
\frac{1}{2}f_{c2} \, .
\end{equation}
The complex-structure moduli $U^i$ are obtained from the conditions to
maintain $\mathcal{N}=1$ supersymmetry.  Consistent with the
complex-structure parameters $\chi_1 \! = \! 2, \, \chi_{2,3} \! = \!
1$, this model has
\begin{equation}
U^1 = 2i~,~~ U^2 = i~,~~ U^3 = i~.~\,
\end{equation}
Note that the conditions for preserving $\mathcal{N}=1$ supersymmetry
do not actually fix one complex structure modulus, $U^3$, and we have 
chosen a value for it which results in gauge coupling unification.  In
a completely realistic compactification, it would be necessary to find
some mechanism to stabilize this modulus, as well as the K\a"ahler and
open-string moduli, and the dilaton.  
In the supergravity basis,
\begin{eqnarray}
\mathrm{Re}\,(s)& =&
\frac{e^{-{\phi}_4}}{2\pi}\,\left(\frac{\sqrt{\mathrm{Im}\,U^{1}\,
\mathrm{Im}\,U^{2}\,\mathrm{Im}\,U^3}}{|U^1U^2U^3|}\right) \, ,
\nonumber \\
\mathrm{Re}\,(u^j)& =&
\frac{e^{-{\phi}_4}}{2\pi}\left(\sqrt{\frac{\mathrm{Im}\,U^{j}}
{\mathrm{Im}\,U^{k}\,\mathrm{Im}\,U^l}}\right)\;
\left|\frac{U^k\,U^l}{U^j}\right| \qquad (j,k,l)=(\overline{1,2,3}) \,
, \nonumber \\
\mathrm{Re}(t^j)&=&\frac{i\alpha'}{T^j} \, , \label{idb:eq:moduli}
\end{eqnarray}
where $\phi_4$ is the four-dimensional dilaton, one finds 
\begin{equation}
\mbox{Re}(s) = \frac{1}{2\pi} \frac{e^{-\phi_4}}{\sqrt{2}} \, ,
\ \ \
\mbox{Re}(u^1) = \frac{1}{2\pi} \frac{e^{-\phi_4}}{\sqrt{2}} \, ,
\ \ \
\mbox{Re}(u^2) = \frac{1}{2\pi} \frac{2e^{-\phi_4}}{\sqrt{2}} \, ,
\ \ \
\mbox{Re}(u^3) = \frac{1}{2\pi} \frac{2e^{-\phi_4}}{\sqrt{2}} \, .
\end{equation}

Inserting these expressions into
Eqs.~(\ref{idb:eq:gkf})--(\ref{kingaugefun}), one finds that the gauge
couplings are unified at the string scale,
\begin{equation}
g^2_{s} = g^2_{w} = \frac{5}{3}g^2_Y = 2\sqrt{2}\pi e^{\phi_4} \, .
\end{equation}
Choosing $\phi_4 = -2.295$ so that $e^{\phi_4}= 0.101$, one obtains
$g^2_{\rm unif} \approx 0.895$, the value of the unified gauge
coupling at GUT scale for a four-generation MSSM\@.  One further finds
the string scale to be
\begin{equation}
M_{St} = \pi^{1/2} e^{\phi_4} M_{\rm Pl} \approx 4.35 \times
10^{17}~\mbox{GeV} \, ,
\end{equation}
where $M_{\rm Pl}$ is the reduced Planck scale, $2.44 \times
10^{18}$~GeV\@.

Fixing the value of $\phi_4$ also fixes the gauge couplings of the
hidden-sector USp(8) groups at $M_{\rm St}$:
\begin{equation}
g_1^2 = g_2^2 = 7.160 \, , \ \ \ \ \ \ g_3^2 = 3.580 \, .
\end{equation}
Using the $\beta$-function coefficients shown in
Table~\ref{MI-Numbers}, the renormalization-group equations (RGEs) for
the USp(8)$_{1,2}$ gauge groups produce strong coupling at a scale
$\Lambda_{1,2} \approx 4.5\times10^{15}$~GeV, while the group
USp(8)$_3$ reaches strong coupling at $\Lambda_3
\approx 8.3\times10^{12}$~GeV\@.  Thus, matter charged under these
groups becomes confined into composite states and decouples not far
below the GUT scale.
  
Thus, at low energies this model is a four-generation MSSM, where the
gauge couplings unify $\approx 2.2\times 10^{16}$~GeV\@.  In addition,
one finds matter charged under the hidden-sector gauge groups that
becomes confined into massive bound states at high-energy scales.
These matter representations all have fractional electric charges,
$Q_{\rm em} = \pm \frac 1 6 , \ \pm \frac 1 2$.  In particular, the
fields charged under USp(8)$_3$ all have half-integer electric charges
and confine into bound states with integer electromagnetic charge at a
scale $O(10^{13})$~GeV, similar to the ``cryptons'' studied
in~\cite{Ellis:1990iu, Benakli:1998ut, Ellis:2004cj, Ellis:2005jc}.
If the lightest of these bound states is electromagnetically neutral
and is metastable with a sufficiently long lifetime, it provides a
good candidate for superheavy cold dark matter (CDM).  This
observation is particularly true for states with masses of order
$\Lambda_3$, since superheavy particles with masses in the range $M =
O(10^{11\mbox{--}13})$~GeV might very well have been produced
naturally through the interaction of the vacuum with the gravitational
field during the post-inflationary reheating phase~\cite{superheavy}.

\section{Yukawa Couplings}

\subsection{Formalism and Construction}

A complete form for the Yukawa couplings arising from D6 branes
wrapping on a full compact space $T^2 \times T^2 \times T^2$ can be
expressed as~\cite{CvePap06,Cremades:2003qj}:
\begin{equation} \label{Yukawas}
Y_{ijk} \propto
\prod_{r=1}^3 \vartheta
\left[\begin{array}{c} \delta^{(r)}\\ \phi^{(r)}
\end{array} \right] (\kappa^{(r)}) \, ,
\end{equation}
where the proportionality constant cancels in mass ratios, and
\begin{equation}
\vartheta \left[\begin{array}{c} \delta^{(r)}\\ \phi^{(r)}
\end{array} \right] (\kappa^{(r)})=\sum_{l \in\Z} e^{\pi
i(\delta^{(r)}+l )^2 \kappa^{(r)}} e^{2\pi i(\delta^{(r)}+l )
\phi^{(r)}},   \label{Dtheta}
\end{equation}
with $r=1,2,3$ denoting the three two-tori.  The input parameters are
given by
\begin{eqnarray} 
\nonumber
&&\delta^{(r)} = \frac{i^{(r)}}{I_{ab}^{(r)}} +
\frac{j^{(r)}}{I_{ca}^{(r)}} + \frac{k^{(r)}}{I_{bc}^{(r)}} +
\frac{d^{(r)} ( I_{ab}^{(r)} \epsilon_c^{(r)} + I_{ca}^{(r)}
\epsilon_b^{(r)} + I_{bc}^{(r)} \epsilon_a^{(r)}
)}{I_{ab}^{(r)} I_{bc}^{(r)} I_{ca}^{(r)}} +
\frac{s^{(r)}}{d^{(r)}}, \\ \nonumber &&\phi^{(r)} =
\frac{I_{bc}^{(r)} \theta_a^{(r)} + I_{ca}^{(r)}
\theta_b^{(r)} + I_{ab}^{(r)} \theta_c^{(r)}}{d^{(r)}}, \\ 
&&\kappa^{(r)} = \frac{J^{(r)}}{\alpha'} \frac{|I_{ab}^{(r)}
I_{bc}^{(r)} I_{ca}^{(r)}|}{(d^{(r)})^2}.
\label{eqn:Yinput}
\end{eqnarray}
where the indices $i^{(r)}$, $j^{(r)}$, and $k^{(r)}$ label the
intersections on the $r^{\rm th}$ torus, $d^{(r)} \! =
gcd(I^{(r)}_{ab}, I^{(r)}_{bc}$, $I^{(r)}_{ca})$, and the integer
$s^{(r)}$ is a function of $i^{(r)}$, $j^{(r)}$, and $k^{(r)}$
corresponding to different ways of counting triplets of intersections.
$J^{(r)}$ is the K\"ahler modulus on the $r^{\rm th}$ torus and
$\alpha^\prime$ is the string tension.  The shift parameters
$\epsilon_a^{(r)}$, $\epsilon_b^{(r)}$, and $\epsilon_c^{(r)}$
correspond to the relative positions of stacks $a$, $b$, and $c$,
while the parameters $\phi_a^{(r)}$, $\phi_b^{(r)}$, and
$\phi_c^{(r)}$ are Wilson lines associated with these stacks.  For
simplicity and to exhibit the flexibility of our model, we set the
Wilson lines to zero.  The brane shifts and Wilson lines together
comprise the open-string moduli, which must be stabilized in a
complete model by some mechanism.  In this paper we treat them as free
parameters; however, we discuss the stabilization of these moduli in
the next subsection.  For simplicity, let us define
\begin{equation}
\epsilon^{(r)} \equiv \frac{d^{(r)} ( I_{ab}^{(r)} \epsilon_c^{(r)} +
I_{ca}^{(r)} \epsilon_b^{(r)} + I_{bc}^{(r)} \epsilon_a^{(r)}
)}{I_{ab}^{(r)} I_{bc}^{(r)} I_{ca}^{(r)}} \, .
\label{TotalBraneShift}
\end{equation}
Although these Yukawa coupling formulas originally refer to $T^6 =
T^2\times T^2 \times T^2$, they may be generalized to $T^6/(\Z_2
\times \Z_2)$ simply by including all of the orbifold images.
However, in the present case the cycles wrapped by the orbifold images
of a stack of D branes $a$ are homologically identical to the original
cycle wrapped by the stack $a$.  In addition, the intersection numbers
between the cycles defined on the orbifold turn out to be the same as
the intersection numbers between those on the ambient torus.  Thus,
the above formulas for $T^6 = T^2\times T^2 \times T^2$ may be used
without change on $T^6/(\Z_2 \times \Z_2)$.
\begin{table}[t]
\footnotesize
\caption{Index assignments labeling the intersections at which each
chiral field is localized.}
\label{tab:indexing}
\begin{center}
\begin{tabular}{|c|c||c|c||c|c|c}
\hline \hline

Left-handed & $i = $ & Right-handed & $j = $ & Higgs field & $k = $ \\
field & $\left\{i^{(1)}, i^{(2)}, i^{(3)}\right\}$&
field & $\left\{j^{(1)}, j^{(2)}, j^{(3)}\right\}$ &
      & $\left\{k^{(1)}, k^{(2)}, k^{(3)}\right\}$ \\

\hline

    $F_L^1(Q^1_L, L^1_L)$& $\left\{0, 0, 0\right\}$ & $F_R^1(Q^1_R,
L^1_R)$ & $\left\{0, 0, 0\right\}$ & $H^1(H^1_u, H^1_d)$ &
$\left\{0, 0, 0\right\}$  \\

    $F_L^2(Q^2_L, L^2_L)$& $\left\{0, 1, 0\right\}$ & $F_R^2(Q^2_R,
L^2_R)$ & $\left\{0, 0, 1\right\}$ & $H^2(H^2_u, H^2_d)$  &
$\left\{1, 0, 0\right\}$   \\

    $F_L^3(Q^3_L, L^3_L)$& $\left\{1, 0, 0\right\}$ & $F_R^3(Q^3_R,
L^3_R)$ & $\left\{1, 0, 0\right\}$ & $H^3(H^3_u, H^3_d)$   &
$\left\{2, 0, 0\right\}$ \\
    
    $F_L^4(Q^4_L, L^4_L)$& $\left\{1, 1, 0\right\}$ & $F_R^4(Q^4_R,
L^4_R)$ & $\left\{1, 0, 1\right\}$ & $H^4(H^4_u, H^4_d)$ &
$\left\{3, 0, 0\right\}$ \\

\hline

\end{tabular}

\end{center}

\end{table}

Since in this model the intersections are distributed over all three
two-tori, they must be carefully labeled.  Let us introduce the
indices $i = \left\{i^{(1)}, i^{(2)}, i^{(3)}\right\}$, $j =
\left\{j^{(1)}, j^{(2)}, j^{(3)}\right\}$, and $k = \left\{k^{(1)},
k^{(2)}, k^{(3)}\right\}$ to label the intersections at which the
left-handed fields, right-handed fields, and Higgs fields,
respectively, are localized.  For this model, these indices can assume
the values
\begin{equation}
\begin{array}{lll}
i^{(1)} = \left\{0, 1\right\}, &
i^{(2)} = \left\{0, 1\right\}, \ \ &
i^{(3)} = \left\{0\right\}, \\
j^{(1)} = \left\{0, 1\right\}, &
j^{(2)} = \left\{0\right\}, &
j^{(3)} = \left\{0, 1\right\}, \\
k^{(1)} = \left\{0, 1, 2, 3\right\}, \ \ &
k^{(2)} = \left\{0\right\}, &
k^{(3)} = \left\{0\right\}, \\
\end{array}
\end{equation}
%
%
as obtained by considering the intersection numbers $I_{ab}^{(r)}$,
$I_{ac}^{(r)}$, $I_{bc}^{(r)}$ on each torus.  Then each left-handed,
right-handed, and Higgs field is identified by a triplet of indices,
as summarized in Table~\ref{tab:indexing}.  The independent Yukawa
couplings $Y^{(r)}_{i^{(r)}j^{(r)}k^{(r)}}$ on each torus are then
labeled:
\begin{eqnarray}
r = 1: & Y^{(1)}_{000}, \ \  Y^{(1)}_{011}, \ \ Y^{(1)}_{101}, \ \
Y^{(1)}_{110}, \ \ Y^{(1)}_{002}, \ \ Y^{(1)}_{013}, \ \
        Y^{(1)}_{103}, \ \ Y^{(1)}_{112}, \\
r = 2: & y_1 \equiv Y^{(2)}_{000}, \ \ y_2 \equiv Y^{(2)}_{100}, \\
r = 3: & z_1 \equiv Y^{(3)}_{000}, \ \ z_2 \equiv Y^{(3)}_{010},  
\end{eqnarray}
being the only remaining components after applying the selection rules
$i^{(r)} + j^{(r)} + k^{(r)} = 0 \! \! \mod d^{(r)}$,
\begin{eqnarray}
i^{(1)} + j^{(1)} + k^{(1)} = 0 \ \mbox{mod} \ 2 \, , \nonumber \\
i^{(2)} + j^{(2)} + k^{(2)} = 0 \ \mbox{mod} \ 1 \, , \\
i^{(3)} + j^{(3)} + k^{(3)} = 0 \ \mbox{mod} \ 1 \, . \nonumber 
\end{eqnarray}
The full Yukawa couplings are then given by a product of the couplings
on each torus,
\begin{equation}
Y_{ijk} = \prod_{r=1}^3 Y^{(r)}_{i^{(r)}j^{(r)}k^{(r)}} \, .
\end{equation}
Thus, the Yukawa matrices (including VEVs) in this model are of the
form
\begin{equation}
\mathcal{Y} \equiv \sum_k Y_{ijk} v_k = \left( \begin{array}{cc}
\left[ Y^{(1)}_{000} v_0 + Y^{(1)}_{002} v_2 \right] Y \! Z &
\left[ Y^{(1)}_{011} v_1 + Y^{(1)}_{013} v_3 \right] Y \! Z \\
\left[ Y^{(1)}_{101} v_1 + Y^{(1)}_{103} v_3 \right] Y \! Z &
\left[ Y^{(1)}_{110} v_0 + Y^{(1)}_{112} v_2 \right] Y \! Z
\end{array} \right) \, ,
\end{equation}
where $Y \! Z$ is the singular 2$\times$2 matrix
\begin{equation}
Y \! Z \equiv \left( \begin{array}{cc}
y_1 z_1 &
y_1 z_2 \\
y_2 z_1 &
y_2 z_2
\end{array} \right) \, , \label{YZ}
\end{equation}
and the values $v_{k^{(1)}} = \left\langle H_{k^{(1)} +1}
\right\rangle$ are the VEVs of the Higgs fields.

Since the same singular submatrix $Y \! Z$ appears in all sub-blocks
of $\mathcal{Y}$, the full Yukawa matrices $\mathcal{Y}$ are only rank
2,
so that at most two eigenvalues of $\mathcal{Y}$ can be different from
zero.  However, this result can be quite desirable, since the third-
and fourth-generation quarks and leptons tend to be significantly
heavier than those of the first and second generations.  In principle,
lifting the rank-2 degeneracies (to induce small masses and mixings)
can be achieved by using higher-order couplings than included here,
such as the fourth-order terms considered
in~\cite{Chen:2008rx,Anastasopoulos:2009mr}, or supersymmetry loop
corrections~\cite{Abel:2003yh}.  For example, the couplings
\begin{eqnarray}
W_4 & \supset y_{ijk} \bar{s}_R^i F_L^j F_R^k H' +
y_{ijk}' \bar{s}_L^i F_L^j F_R^k \bar{H}'.
\end{eqnarray}
can perturb the Yukawa matrices to break the rank degeneracy, or
additional trilinear couplings with the vectorlike Higgs fields $H$
and $\bar{H}'$ can be induced by D-brane instantons.

For this model, the $\delta$ parameters that enter the $\vartheta$
functions of Eq.~(\ref{Dtheta}) are given by
\begin{eqnarray}
\delta^{(1)} &=& \frac{i^{(1)}}{2} - \frac{j^{(1)}}{2} -
\frac{k^{(1)}}{4} - \frac 1 4 ( 2 \epsilon_a^{(1)} + \epsilon_b^{(1)}
- \epsilon_c^{(1)} ) + \frac{s^{(1)}}{2} \, , \\
\delta^{(2)} &=& -\frac{i^{(2)}}{2} + \frac 1 2 ( \epsilon_a^{(2)}
- \epsilon_b^{(2)} - 2 \epsilon_c^{(2)} ) \, , \\
\delta^{(3)} &=& \frac{j^{(3)}}{2} -\frac 1 2 ( \epsilon_a^{(3)}
- 2 \epsilon_b^{(3)} + \epsilon_c^{(3)} ) \, .
\label{thetaparameters}
\end{eqnarray} 
For simplicity, as mentioned above we have taken $\phi^{(1)}=
\phi^{(2)} = \phi^{(3)} = 0$.  Allowed choices for $s^{(1)}$ are
restricted by the requirement that the parameters
$\delta^{(1)}(i,j,k)$ must differ by an integer for triplets $(i,j,k)$
forming triangles with the same area.  In the present case, one finds
two independent choices satisfying this constraint: $s^{(1)}= i^{(1)}$
and $s^{(1)}= j^{(1)}$.  For definiteness, let us make the choice
$s^{(1)}= j^{(1)}$.  We then find
\begin{eqnarray}
Y^{(1)}_{000} = Y^{(1)}_{112} \equiv x_0 \, , & & Y^{(1)}_{011} =
Y^{(1)}_{101} \equiv x_1 \, , \\ \nonumber
Y^{(1)}_{002} = Y^{(1)}_{110} \equiv x_2 \, , & & Y^{(1)}_{013} =
Y^{(1)}_{103} \equiv x_3 \, .
\end{eqnarray}
The Yukawa matrix then takes the simple form
\begin{eqnarray}
\mathcal{Y} = \left(\begin{array}{cc}
( x_0 v_0 + x_2 v_2 ) Y \! Z &
( x_1 v_1 + x_3 v_3 ) Y \! Z \\
( x_1 v_1 + x_3 v_3 ) Y \! Z &
( x_2 v_0 + x_0 v_2 ) Y \! Z
\end{array} \right) \, ,
\end{eqnarray}
where we again see that the Yukawa matrices are rank 2, so that two
eigenvalues are zero.  The four eigenvalues of $\mathcal{Y}$ can be
given in closed form as $\lambda_1 = \lambda_2 = 0$, and the two
nonzero eigenvalues
\begin{eqnarray}
\label{eigenvals2}
\lambda_3 = \frac 1 2 (y_1 z_1 + y_2 z_2)
[(x_0 + x_2)(v_0 - v_2) - \Delta] \, , \\ \nonumber
\lambda_4 = \frac 1 2 (y_1 z_1 + y_2 z_2)
[(x_0 + x_2)(v_0 + v_2) + \Delta] \, ,
\end{eqnarray}
where 
\begin{equation}
\label{Delta}
%
\Delta = \sqrt{(x_0 - x_2)^2 (v_0 - v_2)^2 + 4 (x_1 v_1 + x_3 v_3)^2}
\, ,
\end{equation}
from which one sees that a mass splitting between the third and fourth
generations automatically occurs for generic nonzero values of the
Higgs VEVs.  In order to work with simple closed-form solutions, note
that setting $v_0 = v_2$ simplifies Eq.~(\ref{eigenvals2}) to just
\begin{eqnarray}
\label{eigvalues3}
\lambda_3 = \frac 1 2 (y_1 z_1 + y_2 z_2)\left[(x_0 + x_2)v_0 -
(x_1 v_1 +  x_3 v_3)\right], \\ \nonumber
\lambda_4 = \frac 1 2 (y_1 z_1 + y_2 z_2)\left[(x_0 + x_2)v_0 +
(x_1 v_1 +  x_3 v_3) \right].
\end{eqnarray}
The ratio of the third- to fourth-generation fermion masses is then
given by
\begin{equation}
\frac{m^{f_{u,d}}_{3}}{m^{f_{u,d}}_{4}} =
\left|\frac{x^{f_{u,d}}_0 v^{u,d}_0
- x^{f_{u,d}}_1 v^{u,d}_1
+ x^{f_{u,d}}_2 v^{u,d}_0
- x^{f_{u,d}}_3 v^{u,d}_3}{x^{f_{u,d}}_0 v^{u,d}_0
+ x^{f_{u,d}}_1 v^{u,d}_1 +
x^{f_{u,d}}_2 v^{u,d}_0 + x^{f_{u,d}}_3 v^{u,d}_3}\right| \, ,
\label{massratio}
\end{equation}
where $f_u = \left\{ u, \nu \right\}$, $f_d = \left\{ d, l
\right\}$, couple to the appropriate Higgs VEVs $v^{u,d}_i$,
respectively.

Finally, the eigenstates corresponding to the eigenvalues
Eq.~(\ref{eigenvals2}) or Eq.~(\ref{eigvalues3}) may be expressed as
linear combinations of the fields defined in Table~\ref{tab:indexing},
which are localized at the intersections between D6 branes.  One finds
that the eigenfunctions assume the forms
\begin{eqnarray}
\left|\lambda_3\right\rangle \propto \left( \begin{array}{c}
-\alpha \chi \\
\chi \end{array} \right), \ \ \
\left|\lambda_4\right\rangle \propto \left( \begin{array}{c}
\chi \\
\alpha \chi \end{array} \right), \ \ \ {\rm where} \
\chi = \left( \begin{array}{c} \frac{y_1}{y_2} \\ 1 \end{array}
\right) \, ,
\end{eqnarray}  
where $\alpha \to 1$ in the special case of Eq.~(\ref{eigvalues3}).
Most significantly, the ratios of the third- to fourth-generation
coefficients defining these linear combinations are both given by the
same ratio of the Yukawa couplings on the second two-torus, $y_1/y_2$:
Thus, so long as the Yukawa couplings $y_1$ and $y_2$ are the same for
both the up- and down-type fermions, no mixing occurs between the
third and fourth generations at the trilinear order.  This condition
can be easily satisfied if the D-branes are split only on the first
and third two-tori.\footnote{Likewise, the two massless-state
eigenvectors depend only upon the third-torus parameter combination
$z_1/z_2$.}

\subsection{Analysis}

A full comparison of model predictions to experimental data would
require the full RGE evolution of the observed masses and CKM elements
from the weak scale to the unification scale $M_{\rm GUT}$.  However,
the masses of the fourth-generation fermions, and their mixings with
the other three, are of course unknown~\cite{Nakamura:2010zzi}.  As
discussed above, the combination of direct observation bounds,
electroweak (EW) precision tests, and perturbative unitarity
constraints
place strong constraints on the possible masses, which is just as true
in the SM as in the MSSM~\cite{Dawson:2010jx}.  However,
four-generation models, both supersymmetric and not, are plagued by
the presence of several large Yukawa couplings that exhibit Landau
poles at the TeV scale and above.  While attempts have been made to
stabilize the numerical evolution of the RGEs in four-generation
models up to $M_{\rm GUT}$ by including new matter fields~({\it
e.g.},~\cite{Murdock:2008rx}), questions of the robustness of such
models remain.  For the following analysis, we assume that the
effective running of the Yukawa couplings between the GUT scale and
the EW scale may be stabilized by some mechanism, and we note that
vectorlike matter similar to that introduced in~\cite{Murdock:2008rx}
is present in the model, as seen in Table~\ref{Spectrum}.  For
simplicity, we further (crudely) assume that the running of the Yukawa
couplings between the GUT and EW scales varies slowly, so that
weak-scale values of observables may be used.

Let us recall that the Pati-Salam gauge symmetry has been broken to
the SM by brane splitting.  For the moment, assume that this splitting
occurs only on the first two-torus.  In fact, to forbid this mixing
just for quarks, $c_1 \! = \! c_2$ on the second torus is sufficient.
Then one can write the total brane shifts on the first torus for each
type of fermion, given by Eq.~(\ref{TotalBraneShift}) with $r=1$ as
\begin{eqnarray}
\epsilon^{\nu} = \frac{2\epsilon_{a2} + \epsilon_{b} -
\epsilon_{c1}}{4}\, , \ \ \ \ \ \ \ \ \ \epsilon^{u} =
\frac{2\epsilon_{a1} + \epsilon_{b} - \epsilon_{c1}}{4}\, , \\
\nonumber 
\epsilon^{l} = \frac{2\epsilon_{a2} + \epsilon_{b} -
\epsilon_{c2}}{4}\, , \ \ \ \ \ \ \ \ \ \epsilon^{d} =
\frac{2\epsilon_{a1} + \epsilon_{b} - \epsilon_{c2}}{4}\, ,
\end{eqnarray}
where quarks begin on stack $a1$ and leptons on stack $a2$, while the
neutrinos and up-type quarks both end on stack $c1$, and the charged
leptons and down-type quarks both end on stack $c2$. 
One then observes that
\begin{eqnarray}
\epsilon^{u} - \epsilon^{d} = \frac{1}{4}(-\epsilon_{c1} +
\epsilon_{c2}) = \epsilon^{\nu} - \epsilon^{l}.
\label{shiftconstraint1} 
\end{eqnarray}
Taking, for example, $\epsilon^{\nu}=0$, one must then satisfy the
constraint
\begin{equation}
\epsilon^{d} = \epsilon^{u} + \epsilon^{l},
\label{shiftdiff}
\end{equation} 
when choosing parameters to fit the mass ratios of third- and
fourth-generation fermions.  Note that although we are treating the
brane-shifts as free parameters, these are open-string moduli that
should be stabilized in a completely realistic compactification.  This
requirement can be satisfied if the D-branes wrap rigid cycles, which
are available on several different backgrounds~\cite{Forste:2010gw,
Dudas:2005jx, Blumenhagen:2005tn}. In addition, the K\a"ahler moduli
on each torus may be stabilized by including supergravity fluxes
and/or by gaugino condensation in a hidden
sector~\cite{BluCveLanShi05}.

\begin{figure}[t]
\centering
\includegraphics{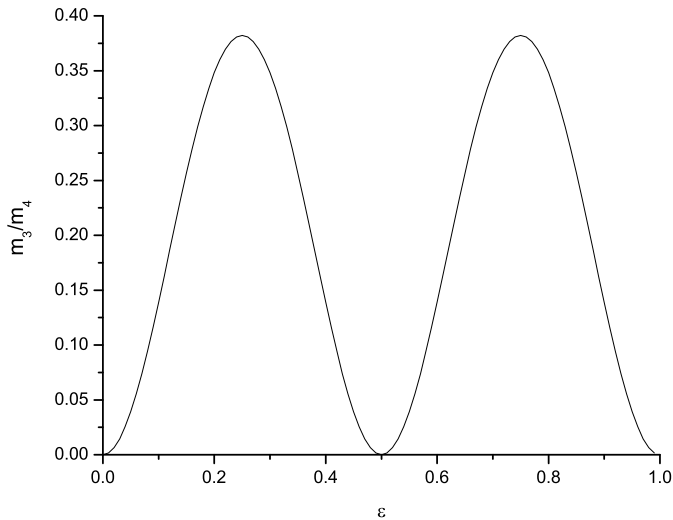}
\caption{Ratio of the $t$ to $t'$ quark masses as a function
of the shift parameter $\epsilon \equiv \epsilon^{u}$, where the
Higgs VEVs are chosen to satisfy Eq.~(\ref{UpHiggsVEVs}), and $\kappa
= \sqrt{3} \pi$.}
\label{fig:MassRatio34U}
\end{figure}

From the relations given in the previous subsection, we make certain
general observations.  First, Eq.~(\ref{massratio}) shows that a very
light third-generation neutrino $\nu_{\tau}$ can readily be obtained
if
\begin{equation}
(x_0^{\nu} + x_2^{\nu})v_0^u -  x_1^{\nu} v_1^u -  x_3^{\nu} v_3^u
\approx 0.
\end{equation}
where we take $v_0 = v_2$ so that the generic eigenvalues of the
Yukawa matrices are given by Eq.~(\ref{eigvalues3}).  This condition
can easily be satisfied if, for example, the Higgs VEVs take the
values
\begin{equation}
v_0^u = 2C/(x_{0}^{\nu}+x_{2}^{\nu}), \ \ \ \ \ v_1^u = C/x_{1}^{\nu},
\ \ \ \ \ v_3^u = C/x_{3}^{\nu}
\label{UpHiggsVEVs}
\end{equation}
where $C$ is constant.  Using these values,
Eq.~(\ref{eigvalues3}) shows that $m_{\nu_{\tau}}$ vanishes while
$m_{\nu^\prime}$ may be large.  With these VEV choices, the neutrino
Yukawa matrix takes the form
\begin{eqnarray}
\mathcal{Y}^{\nu} \sim 2C \left( \begin{array}{cc}
Y \! Z & Y \! Z \\
Y \! Z & Y \! Z
\end{array} \right) \, ,
\end{eqnarray}
and essentially degenerates to rank 1, so that only $\nu^\prime$
receives mass at leading order.  Note that this rank-1 condition is
independent of the specific values assigned to the K\"ahler modulus
and shift parameter $\epsilon^{\nu}$.

Since the up-type quark masses depend upon the same Higgs VEVs as the
neutrinos, the question now becomes whether it is possible to obtain
masses for the $t$ and $t'$ quarks in the range
\begin{equation}
\frac{m_{t'}}{m_t} \approx \mbox{2--3} \, .
\label{43upquarkratio}
\end{equation}
At first glance, simultaneously satisfying these criteria may appear
difficult.  However, the Yukawa couplings $x_i^{u}$ are in general
different than $x_i^\nu$ since the position on the first two-torus of
the $a1$ stack of branes upon which the quarks begin may be different
than the position the $a2$ stack from which the leptons begin, so that
$\epsilon_{a1} \neq \epsilon_{a2}$, and $\epsilon^{u} \! - \!
\epsilon^\nu \! = \! \frac 1 2 ( \epsilon_{a1} \! - \! \epsilon_{a2})$.

We now show that one can obtain results consistent with
Eq.~(\ref{43upquarkratio}) for generic nonzero values of $\kappa \!
\equiv \! -i\kappa^{(1)}$ and $\epsilon^{u}$.  For example, let us
choose $\epsilon^{\nu} = 0$ and $\kappa \! = \! \sqrt{3}\pi \!$
(an arbitrary value used purely for illustration), while the
Higgs VEVs have the values given by Eq.~(\ref{UpHiggsVEVs}).  The
resulting ratio $m_t/m_{t'}$ given by Eq.~(\ref{UpHiggsVEVs}) as a
function of the shift parameter $\epsilon \! \equiv \! \epsilon^{u}$
is plotted in Fig.~\ref{fig:MassRatio34U}.  This plot shows that mass
ratios satisfying Eq.~(\ref{43upquarkratio}) can be obtained for
\begin{equation}
\epsilon^{u} \approx \mbox{0.19--0.31}, \ \ \ \ \ {\rm and}
\ \ \ \ \ \epsilon^{u} \approx \mbox{0.69--0.81}\  \, .
\end{equation}
\begin{figure}[t]
\centering
\includegraphics{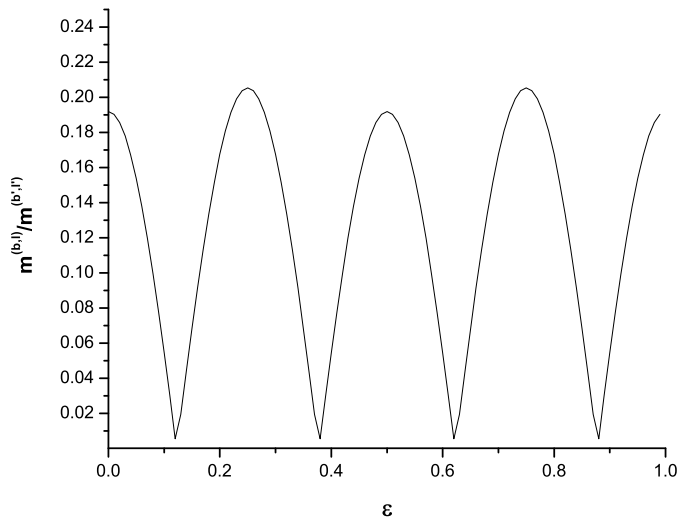}
\caption{Mass ratio between the third- and fourth-generation down-type
quark (and charged-lepton masses) as functions of K\"ahler modulus
$\kappa \! = \! \sqrt{3}\pi$ on the first torus and shift parameter
$\epsilon \! \equiv \epsilon^{d}$.}
\label{fig:MassRatio34DL}
\end{figure}

Similarly, the down-type quark masses and charged leptons depend upon
the Higgs VEVs $v_k^d$.  However, in addition the shift parameters
must satisfy Eq.~(\ref{shiftdiff}).
%
%
The Higgs VEVs $v_k^d$ may be chosen freely, apart from overall
normalization; for simplicity, let us choose
\begin{equation}
v_0^d = 2D/({x'}_{0}^{\nu}+{x'}_{2}^{\nu}), \ \ \ \ \
v_1^d = D/{x'}_{1}^{\nu}, \ \ \ \ \ v_3^d = D/{x'}_{3}^{\nu},
\label{DownHiggsVEVs}
\end{equation}
where $D$ is a constant and ${x'}_{k}^{\nu}$ are the same $\vartheta$
functions used in calculating the eigenvalues in the neutrino sector,
but with a different value of the input parameter $\kappa$, which we
label $\kappa^\prime$.  Note that the Yukawa couplings for the
down-type quarks and charged leptons must be calculated using the same
K\"ahler modulus $\kappa = \sqrt{3}\pi$ as used for the up-type
quarks and neutrinos.  However, the down-type Higgs VEVs are chosen by
the ansatz Eq.~(\ref{DownHiggsVEVs}), with $\kappa^\prime < \kappa$.
Obviously, many other suitable choices are possible.

In Fig.~\ref{fig:MassRatio34DL} we plot the ratio of third- to
fourth-generation masses for the down-type quarks and charged leptons
using $\kappa = \sqrt{3}\pi$ and $\kappa^\prime =\sqrt{2}\pi/2$
(again, purely an illustrative choice), as a function of brane shift
parameter $\epsilon \! \equiv \! \epsilon_d$.  As one sees, the
resulting function exhibits several maxima, with minima that are
equally spaced about the maxima.  The appearance of these alternating
maxima and minima can, of course, be traced to the fact that the mass
eigenvalues are essentially linear superpositions of the $\vartheta$
functions exhibiting well-defined symmetry properties in $\epsilon$.

In fitting the third- to fourth-generation mass ratios for the
down-type quarks and charged leptons, we require
\begin{equation}
\frac{m_{b'}}{m_{b}} \approx \mbox{80--130} \, , \ \ \ \ \ \ \ \ \ \ \
\ \frac{m_{\tau'}}{m_{\tau}} \approx \mbox{60--700} \, ,
\end{equation}  
These mass ratios are clearly much larger than $m_{t'}/m_t$.
Furthermore, one must impose Eq.~(\ref{shiftdiff}) on the shift
parameters.  Fortunately, as noted previously, the mass ratio between
third- and fourth-generation down-type quarks and neutrinos plotted in
Fig.~\ref{fig:MassRatio34DL} exhibit minima near $\epsilon = 0.375$,
$\epsilon = 0.125$, which are separated by $\Delta \epsilon \simeq
0.25$.  By choosing values near $\epsilon^{d} \approx 0.375$ and
$\epsilon^l \approx 0.625$, satisfying all constraints is possible.
Indeed, if one chooses $\epsilon^{d}=0.3737$ and $\epsilon^l =
0.1237$, for example, one finds
\begin{equation}
\frac{m_{b'}}{m_{b}} = 97.7 \ \ \ \ \ \mbox{and} \ \ \ \ \
\frac{m_{\tau'}}{m_{\tau}} = 267.3 \, .
\end{equation}
Choosing nearly symmetric shift values naturally provides nearly equal
Yukawa couplings for the $b$ and $\tau$, so that $b$-$\tau$
unification may be naturally achieved.  Since $\epsilon^{u} \! = \!
\epsilon^{d} - \epsilon^{l} = 0.25$, one obtains a value of
$\epsilon^{u}$ near a maximum of Fig~\ref{fig:MassRatio34U}.  One then
obtains the phenomenologically acceptable value
\begin{equation}
\frac{m_{t'}}{m_t} = 2.62 \, .
\end{equation}
Furthermore, noting that
\begin{equation}
\frac{m_{\nu^{\, \prime}}}{m_{t'}} =
\left| \frac{x^{\nu}_0 v^{u}_0 + x^{\nu}_1 v^{u}_1
+ x^{\nu}_2 v^{u}_0 + x^{\nu}_3 v^{u}_3}
{x^{u}_0 v^{u}_0 + x^{u}_1 v^{u}_1 +
x^{u}_2 v^{u}_0 + x^{u}_3 v^{u}_3} \right| \, ,
\label{massratio2}
\end{equation}
where $x^{\nu}_i$ and $x^{u}_i$ differ only by the use of the shift
parameters $\epsilon^{\nu} \! = \! 0$ and $\epsilon^u \! = \!  0.25$,
respectively, one finds
\begin{equation}
\frac{m_{\nu^{\, \prime}}}{m_{t'}} = 0.924 \, .
\end{equation}
Using the tabulated masses $m_{t, b, \tau}$~\cite{Nakamura:2010zzi}
and these ratios, one obtains $m_{t^\prime} \! = \! 452.5$~GeV,
$m_{b'} \! = \! 409.5$~GeV $\rightarrow m_{t'} \! - \! m_{b'} \! = \!
43.0$~GeV, and $m_{\tau^\prime} \! = \! 474.9$~GeV, $m_{\nu^{\,
\prime}} \! = \!  418.1$~GeV $\rightarrow m_{\tau^\prime} \! - \!
m_{\nu^{\, \prime}} \!  = \!  56.8$~GeV, meaning that all the mass
constraints listed in the Introduction are satisfied.

In addition, no mixing occurs between the third- and fourth-generation
fermions since, as discussed above, the D-branes have only been split
on the first two-torus.  We also note that one can split the D-branes
on the third two-torus as well, without inducing mixing between the
third and fourth generations.  Doing so does not affect the mass
ratios $m_3^f/m_4^f$ for each type $f$ of fermion.  However, splitting
on the third two-torus does affect the relative mass hierarchy between
the up-type quarks and the neutrinos, as well as the between the
down-type quarks and charged leptons.  The suppression of
fourth-generation quark mixing implies long quark lifetimes and the
possibility of hadronic bound states~\cite{LongFourLife}, as discussed
in~\cite{Ishiwata:2011ny,Hung:bound}.

As seen in this analysis, one can obtain mass splittings within each
generation in a fairly natural way, since the mass ratios involve
superpositions of $\vartheta$ functions that have convenient symmetry
properties.  As a result, the mass ratios between the third- and
fourth-generation fermions exhibit maxima and minima as a function of
shift parameter $\epsilon$.  Note that generic points reflect fully
rank-2 Yukawa matrices, while the minima correspond [according to
Eq.~(\ref{massratio})] to points where the Yukawa matrices degenerate
to rank 1.  This tendency toward Yukawa matrices of lower rank was
also observed in the rank-4 model of~\cite{Belitsky:2010zr}.  The mass
ratio $m_t/m_{t'}$ is constrained to be much larger than the ratio
between $m_b/m_{b'}$, $m_\tau/m_{\tau^\prime}$, and especially
$m_{\nu_{\tau}}/m_{\nu^{\, \prime}}$.  However, one sees that these
ratios can be obtained if the shift parameter $\epsilon^{u}$ is chosen
to lie near a maximum of the mass ratio curve, while the shift
parameters $\epsilon^{d}$, $\epsilon^{l}$, and $\epsilon^{\nu}$
correspond to minima.  For the present model, the brane shift
parameters $\epsilon^{u}$, $\epsilon^{d}$, $\epsilon^{l}$, and
$\epsilon^{\nu}$ are continuous parameters constrained by
Eq.~(\ref{shiftconstraint1}).

\section{Conclusion}

We have presented a globally consistent four-family MSSM constructed
from intersecting D6 branes wrapping cycles on a $T^6/(\Z_2 \times
\Z_2)$ orientifold.  In contrast to other similar constructions, the
Higgs fields appear in the chiral sector rather than as vectorlike
matter that occurs between nonintersecting D branes.  We find the
gauge couplings to be unified at the string scale and the
hidden-sector gauge groups to become confining in the range
10$^{13}$--10$^{16}$~GeV\@.  Thus, the model is a four-generation MSSM
at low energies.

The Yukawa matrices of the model are rank 2, so that only the third-
and fourth-generation fermions receive masses at trilinear order.  The
first two generations, in principle, receive masses via higher-order
corrections, and so should be naturally lighter.  This result
contrasts sharply with that of our earlier rank-4
model~\cite{Belitsky:2010zr}, in which only small regions of the VEV
parameter space accommodate the known fermion mass and mixing
spectrum.  In addition, the third- and fourth-generation fermions have
nondegenerate masses but do not mix at leading order.  Finally, a
numerical analysis of the mass ratios between the third and fourth
generations shows that simple, elegant, and fairly natural solutions
satisfying all constraints are relatively easy to find.  In
particular, we have shown that one can obtain three light neutrinos
and a heavy fourth, while simultaneously obtaining heavy $t$ and $t'$
quarks in a completely natural way.  In addition, one may easily
obtain masses for the $b'$ quark and $\tau'$ lepton that are heavy in
comparison to the $b$ quark and $\tau$ lepton, respectively.  Moreover,
$b$-$\tau$ unification may be naturally realized.  We conclude that
our analysis demonstrates that the putative fourth-generation fermion
masses required to satisfy the constraints placed upon them by direct
experimental observations, precision electroweak measurements, and
perturbative unitarity constraints can emerge in a completely natural
fashion.  The very real possibility that a fourth generation of chiral
fermions exists deserves serious consideration, one that will be
settled conclusively at the LHC.
   
\section{Acknowledgments}

This work was supported by the National Science Foundation
under Grant No.\ PHY-0757394.

\end{document}